\documentclass[12pt,a4paper]{article}

\usepackage[british]{babel}

\usepackage[a4paper,top=2cm,bottom=2cm,left=2.5cm,right=2.5cm,marginparwidth=1.75cm]{geometry}

\usepackage[style=apa, backend=biber]{biblatex} 
\usepackage{graphicx}
\usepackage[misc]{ifsym}
\usepackage{booktabs}
\usepackage{multirow}
\usepackage{enumitem}
\usepackage{siunitx}

\newcommand*{\SuperScriptSameStyle}[1]{%
  \ensuremath{%
    \mathchoice
      {{}^{\displaystyle #1}}%
      {{}^{\textstyle #1}}%
      {{}^{\scriptstyle #1}}%
      {{}^{\scriptscriptstyle #1}}%
  }%
}
\newcommand*{\oneS}{\SuperScriptSameStyle{*}}
\newcommand*{\twoS}{\SuperScriptSameStyle{**}}
\newcommand*{\threeS}{\SuperScriptSameStyle{*{*}*}}
\newcommand*{\fourS}{\SuperScriptSameStyle{*{**}*}}

\DeclareLanguageMapping{british}{british-apa} 
\DeclareFieldFormat[article]{volume}{\apanum{#1}} 

\usepackage{amsmath}
\usepackage{graphicx}
\usepackage[colorlinks=true, allcolors=blue]{hyperref}
\usepackage{hyperref}
\usepackage{orcidlink}
\usepackage[title]{appendix}
\usepackage{mathrsfs}
\usepackage{amsfonts}
\usepackage{booktabs} 
\usepackage{caption}  
\usepackage{threeparttable} 
\usepackage{algorithm}
\usepackage{algorithmicx}
\usepackage{algpseudocode}
\usepackage{listings}
\usepackage{enumitem}
\usepackage{chngcntr}
\usepackage{booktabs}
\usepackage{lipsum}
\usepackage{subcaption}
\usepackage{authblk}
\usepackage[T1]{fontenc}    
\usepackage{csquotes}       
\usepackage{diagbox}

\usepackage{setspace}
\usepackage{titlesec}
\titleformat{\section} 
  {\normalfont\Large\bfseries}{\thesection.}{1em}{}
  
\usepackage{lineno} 

\rightlinenumbers 

\usepackage{float}   
\usepackage{caption} 


\title{Ink and Algorithm: Exploring Temporal Dynamics in Human-AI Collaborative Writing}

\author[1]{Kaixun Yang}
\author[1]{Yixin Cheng}
\author[1]{Linxuan Zhao}
\author[1]{Mladen Rakovi\'{c}}
\author[1]{Zachari Swiecki}
\author[1]{Dragan Ga\v{s}evi\'{c}}
\author[1,*]{Guanliang Chen}

\affil[1]{Centre for Learning Analytics, Monash University, Melbourne, Victoria, Australia}
\affil[*]{Corresponding author: \texttt{Guanliang.Chen@monash.edu}}

\date{}

\begin{document}
\maketitle

\begin{abstract}
The advent of Generative Artificial Intelligence (GAI) has revolutionized the field of writing, marking a shift towards human-AI collaborative writing in education. However, the dynamics of human-AI interaction in the collaborative writing process are not well understood, and thus it remains largely unknown how human learning can be effectively supported with such cutting-edge GAI technologies. In this study, we aim to bridge this gap by investigating how humans employ GAI in collaborative writing and examining the interplay between the patterns of GAI usage and human writing behaviors. Considering the potential varying degrees to which people rely on GAI usage, we proposed to use Dynamic Time Warping time-series clustering for the identification and analysis of common temporal patterns in AI usage during the human-AI collaborative writing processes. Additionally, we incorporated Epistemic Network Analysis to reveal the correlation between GAI usage and human writing behaviors that reflect cognitive processes (i.e., knowledge telling, knowledge transformation, and cognitive presence), aiming to offer insights for developing better approaches and tools to support human to learn effectively via such human-AI collaborative writing activities. Our findings reveal four major distinct temporal patterns in AI utilization and highlight significant correlations between these patterns and human writing behaviors. These findings have significant implications for effectively supporting human learning with GAI in educational writing tasks.
\end{abstract}

\textbf{Keywords}: Human-AI Collaborative Writing, Time-series Clustering, Generative AI, Epistemic Network Analysis

\section{Introduction} \label{sec:introduction}
Writing serves as a crucial medium for expressing and sharing our thoughts, emotions, and perspectives with others \parencite{dj2015using}. Moreover, the essence of writing extends beyond merely generating ideas; it involves the adept composition and structuring of writing elements \parencite{setyowati2016analyzing}. Consequently, challenges in writing, including issues with vocabulary, grammatical accuracy, and the organization of ideas, can obstruct clear and effective communication \parencite{fareed2016esl}. To address these challenges, both academic and industrial sectors have begun developing online writing tools to support writing based on computational linguistics and natural language processing. These tools contain machine translation platforms such as DeepL, digital writing assistants like Grammarly, and automated paraphrasing applications such as QuillBot \parencite{roe2023review}. Given the strong language comprehension and contextual understanding demonstrated by Large Language Models (LLMs) \parencite{chang2023survey}, recent research endeavors have pivoted towards utilizing these LLMs to enhance AI-powered writing support \parencite{zhao2022leveraging}. 

One such tool is CoAuthor, a human-AI collaborative writing instrument based on GPT-3 \parencite{lee2022coauthor}. This tool provides writers with a variety of writing prompts (either argumentative writing prompts or creative writing prompts) and inspires them to conduct essay composition in response to these prompts. Writers can seek suggestions from GAI and receive up to five sentence suggestions for each request at any point during their writing process. These AI-generated proposals are subject to the writer's discretion, allowing for acceptance, rejection, or modification. Designed primarily as an experimental platform to explore the dynamics of human-AI interaction in the context of writing, CoAuthor not only preserves the final written product but also click-stream events throughout the entire essay composition process.

With these efforts made in developing human-AI collaborative writing systems and collecting human-AI writing datasets \parencite{coenen2021wordcraft, kreminski2020we}, researchers have started investigating the human writing behaviors in this particular setting with the goal of effectively supporting human-AI writing and enhancing human learning \parencite{cheng2024evidencecentered}. Current research often categorizes writers based on the frequency of their AI usage at a rather coarse level throughout the whole writing session (e.g., high vs. low AI usage) \parencite{shibani2023visual}, which sheds little light on the temporal dynamics between human-AI interactions in the collaborative writing process. For instance, one writer might use AI extensively at the start of their writing and then gradually decrease its use; another might initially write without AI assistance but increasingly rely on it as the process progresses. Although these two writers might exhibit similar overall AI usage when evaluated based on the final product (i.e., the total number of times AI suggestions are sought), their temporal AI usage patterns are distinctly different. Additionally, how these AI usage patterns interact with human writing behaviors and their subsequent impact on writing performance and human learning remain largely unexplored. Understanding these interactions could offer valuable insights for educators in evaluating the writing processes of writers with different AI usage patterns and developing more tailored and effective writing tools.

Hence, the study reported in the paper sought to (i) explore common temporal patterns in human interaction with GAI in collaborative writing tasks and (ii) examine how these patterns reflect human cognitive behaviors in writing. Such an investigation offers insights for developing better approaches and tools that cater to the diverse cognitive needs to support human learning effectively through human-AI collaborative writing activities. Formally, this study was guided by the following two \textbf{R}esearch \textbf{Q}uestions:

\begin{enumerate}[label=\bfseries RQ\arabic*,leftmargin = 30pt]
    \item What are the common temporal GAI-usage patterns during human-AI collaborative writing tasks?
    \item To what extent do human writing behaviors correlate with their usage of GAI?
\end{enumerate}

To answer the RQs, we chose a public dataset consisting of 1,445 writing sessions that took place using the CoAuthor tool. The dataset contains not only the final written products, but also log trace events recorded throughout the entire writing process. Generally, writing sessions vary in length, containing different amounts of keystroke logs. Standard clustering techniques, like K-means, are not well-equipped to manage datasets with such variations. For this reason, to answer RQ1, we opted for Dynamic Time Warping (DTW) \parencite{muller2007dynamic}, a method capable of assessing the similarity between two action sequences irrespective of their length disparities. Additionally, to uncover the potential impact of GAI usage on human writing behaviors that reflect cognitive processes (RQ2), we conducted Epistemic Network Analysis (ENA) \parencite{shaffer2017epistemic}, a popular learning analytics method that utilizes network models to depict epistemic actions, on each cluster to evidence the cognitive behaviors inherent to co-writing with GAI. The details are provided in Section \ref{sec:method}.

\section{Related Work} \label{sec:relatedwork}
\subsection{Writing Process Analysis}
Since the past decade or two, analyzing the writing process was deemed challenging due to the difficulties in tracking activities in handwritten form \parencite{sinharay2019prediction}. However, the advent of digital writing tools has made it more feasible to observe and reconstruct the writing process using keystroke logging. Keystroke logging is a prevalent method for examining the writing process in educational settings, involving the recording and timestamping of keystroke actions to reconstruct the writing process \parencite{leijten2013keystroke}. 

Current research primarily centers on deriving writing behaviours, such as between-word pauses, the number of writing bursts, and the number of deleted characters, from keystroke logging. These behaviors are then used to explore their relationships with other writing aspects such as essay quality \parencite{Vakkari2021PredictingEQ}, task complexity \parencite{revesz2017effects}, and language proficiency \parencite{anak2012relationship}. Additionally, investigating the cognitive processes behind writing behaviours is also a significant area of interest in existing research. Writing encompasses a set of recursive and intertwined cognitive processes, e.g., including planning, translating, reviewing, and monitoring \parencite{koppenhaver2010conceptual}. Therefore, it is essential to understand the patterns in writer's cognitive processes to better support them in a writing task \parencite{zhang2015process}. \cite{latif2021remodeling} proposed a writing process model composed of seven components (e.g., Linguistic rehearsing, Reviewing, and Ideational planning), and offered explanations of the strategies used in each component. \cite{conijn2019understanding} proposed the method of mapping features from the keystroke logs to higher-level cognitive processes, such as planning and revising. \cite{baaijen2012keystroke} developed methods and measures for analyzing keystroke logging with the goal of enhancing the correlation between keystroke data and cognitive processes.

Nevertheless, the advent of human-AI collaborative writing is altering the conventional cognitive processes in writing, as AI assumes part of the cognitive burden \parencite{knowles2022human}. Moreover, this collaboration has introduced new writing behaviours, such as the frequency of requesting suggestions from AI, accepting these suggestions, and rejecting them, which were not present in traditional human-only writing scenarios. Consequently, there is a need for further investigation into the cognitive process in human-AI collaborative writing.

\subsection{Study on Human-AI Collaborative Writing}
Existing research on human-AI collaborative writing is relatively scarce. \cite{shibani2023visual} introduced CoAuthorViz, a tool for visualizing keystroke logs in such collaborations. This tool demonstrates various human writing behaviors when assisted by GPT-3 and develops metrics for evaluating the adoption of GPT-3 suggestions in relation to writing quality. \cite{wambsganss2023unraveling} investigated the transmission of bias in a human-AI collaborative writing pipeline. Their study assessed gender bias in essays among various groups, such as those using fine-tuned GPT-2 compared to GPT-3 models. \cite{cheng2024evidencecentered} suggested a methodology based on learning analytics for evaluating human-AI collaborative writing. Their approach involves comparing outcomes across different groups (i.e., User vs. GAI ownership, Creative vs. Argumentative writing, and High vs. Low GPT Temperature settings). \cite{nguyen2024human} suggested using the Hidden Markov Model (HMM) alongside hierarchical sequence clustering for conducting sequence analysis of human-AI interactions in academic writing.

However, these studies often have two primary shortcomings: (i) they generally categorized writers based on their overall AI usage throughout the entire writing session, overlooking the temporal dynamics of these behaviors as the progression of a writing session (studies by Shibani et al., Wambsganss et al., and Cheng et al.); (ii) they revealed limited insight into writers' writing behaviors and the corresponding cognitive processes and thus offered limited implications for supporting human learning in this particular task (studies by Shibani et al., Wambsganss et al., and Nguyen et al.).

Consistent with \cite{cheng2024evidencecentered}, we adopted an evidence-centered assessment framework for evaluating the human-AI collaborative writing processes. Our extension of their work includes the consideration of temporal information in AI usage to characterise different human-AI interaction patterns. Their framework correlates writing behaviors with three cognitive processes:

\begin{itemize}
 \item \texttt{Knowledge Telling}: This is a straightforward writing cognitive process where individuals primarily focus on writing down information as they remember it, without significant engagement or critical reorganization \parencite{bereiter2013psychology}. This behavior is observed when learners accept AI-generated suggestions without making any significant changes, essentially following the AI's guidance as it is presented.
 \item \texttt{Knowledge Transformation}: This involves a more engaged approach, where writers thoughtfully restructure their arguments, integrate various viewpoints, and craft narratives that are both cohesive and impactful \parencite{bereiter2013psychology}. In this context, knowledge transformation is evident when users use AI suggestions as a starting point for revision, adapting them to suit their own needs.
 \item \texttt{Cognitive Presence}: This concept pertains to the degree to which learners can construct and validate meaning through communication within a specific sociocultural context \parencite{garrison1999critical}. It unfolds through a four-stage cycle: triggering event (i.e., the initial encounter with a problem or question that ignites curiosity), exploration (i.e., learners actively seek information to deepen their understanding to the question), integration (i.e., gathered information is synthesized to formulate coherent responses), and resolution (i.e., synthesized knowledge is used to propose a solution). In this context, triggering events might occur when writers approach the AI with a writing-related question or problem. Exploration happens as they engage with the AI to brainstorm or consider different suggestions, exploring various perspectives or writing styles. Integration is the process of synthesizing the ideas and suggestions from the AI to create a unified piece of writing. Finally, resolution occurs when writers refine their work with the AI's help, seeking input on improving language and structure to satisfactorily address their initial problem.
\end{itemize}

\subsection{Time-Series Clustering in Education}
Clustering is a data mining technique that involves grouping similar data into related or homogeneous cluster without prior knowledge of the definitions of these groups \parencite{rai2010survey}. A specific type of clustering is time-series clustering, which deals with sequences of continuous, real-valued elements, known as time-series. Time-series data is commonly encountered in daily life, such as stock market prices, weather statistics, and biomedical records. Methods for clustering time-series data are commonly divided into three classes \parencite{aghabozorgi2015time}: model-based, feature-based, and shape-based approaches. In the field of education, the majority of research has utilized feature-based \parencite{talebinamvar2022clustering, zhang2019identifying, hassan2022utilizing} and shape-based \parencite{shen2017clustering, ma2022grading, he2023clustering} methods.

In \textbf{model-based} time-series approaches, each time-series data is represented using statistical models like HMM. Subsequently, a similarity metric is calculated between these models' parameters. Finally, time-series data are grouped into clusters based on these similarity measures. To the best of our knowledge, there are limited publications that employ model-based time-series clustering in the field of education.

In \textbf{feature-based} time-series approaches, time-series data is converted into a collection of features that capture key characteristics of the data (e.g., mean, variance, and frequency). Clustering is then executed based on these extracted features. Feature-based approaches are the most frequently utilized methods for clustering in education. \cite{talebinamvar2022clustering} processed keystroke log data from student essays, extracting 14 process indicators (e.g., Pause variance, Total keystrokes, and Mean word length) for cluster identification within the dataset. Similarly, \cite{zhang2019identifying} analyzed keystroke logs from student essays, focusing on four fundamental performance indicators (i.e., time spent on task, essay score, number of keystrokes, and typing speed) to detect clusters in the data. In addition to analyzing student writing, feature-based methods are utilized in various other educational areas, including predicting students' academic performance \parencite{hassan2022utilizing}, identifying at-risk students \parencite{hung2015identifying}, and modeling student participation in online discussions \parencite{cobo2010modeling}. 

In \textbf{shape-based} time-series approaches, the focus is on grouping time-series data based on the shape of their curves, rather than just considering the numerical values at particular time points. Dynamic Time Warping (DTW) \parencite{muller2007dynamic} is typically employed as the similarity measure in shape-based approaches. Shape-based approaches have also been implemented in some educational contexts. \cite{shen2017clustering} grouped students according to their moment-by-moment interactive trajectories with educational systems. \cite{ma2022grading} utilized DTW clustering to categorize students based on their historical academic performance. \cite{he2023clustering} clustered students based on their navigational patterns in digital reading tasks considering the sequence of page transitions and time spent on each page.

While time-series clustering is extensively applied in educational settings, the majority of these applications concentrate on feature-based methods. Research exploring the use of shape-based time-series clustering in human-AI collaborative writing tasks is notably absent. Our work not only pioneers in this area but also makes a significant contribution to future studies in this field.

\section{Methodology} \label{sec:method}
\subsection{Dataset}
The CoAuthor dataset \footnote{\url{https://coauthor.stanford.edu/}} is a publicly available dataset collected using the CoAuthor writing tool detailed in Section \ref{sec:introduction}. This dataset was created by 63 writers recruited from the Amazon Mechanical Turk platform. The collection process included a "qualification round" aimed at verifying that participants were capable of (i) knowing how to collaboratively write with AI and (ii) writing a short, interesting story that has a clear ending. Each writer participated in multiple sessions, contributing to a collection of 20 writing prompts. Overall, the dataset comprises 1,445 writing sessions, split into 830 creative writing and 615 argumentative writing sessions. It offers a detailed log of keystroke-level interactions during these sessions, capturing 13 different event types (e.g., text-insertion, text-deletion, and suggestion-get). Each event is characterized by 17 pieces of information (e.g., event name, event source, and timestamp). Additionally, the dataset includes survey data \footnote{\url{https://docs.google.com/spreadsheets/d/1O3EXJm52TQHfFSbzVGZmNIzzdu5ow6IjnOBrGTUY02o}} consisted of five categories: writer information (e.g., native or non-native English speakers), benefits of collaborative writing (i.e., the known benefits of human-human collaborative writing), perceived capabilities of Language Models, perceived limitations of Language Models, and overall experiences. The survey responses are scored on a scale from 1 to 7, indicating a range from strongly disagree to strongly agree. The original paper \cite{lee2022coauthor} provides an in-depth description of the dataset. Given that the dataset provides comprehensive and detailed time-series information on human collaborative writing with AI, we based our experiments on this dataset.

\subsection{RQ1: Dynamic Time Warping Time-Series Clustering}

\noindent\textbf{Dynamic Time Warping (DTW).} The DTW algorithm measures similarity between time-series data by minimizing the effects of shifting and distortion in time by allowing 'elastic' transformation of the time-series in order to detect similar shapes with different phases \parencite{senin2008dynamic}. Consider two time-series sequences $X = \{x_1, x_2, ..., x_n\}$ and $Y = \{y_1, y_2, ..., y_m\}$, where $n$ and $m$ represent the respective lengths of these sequences, and it is possible that $n$ is not equal to $m$. DTW finds the optimal alignment between two sequences, with the similarity score represented by the cost of the optimal warping path. To identify this path, an $ n \times m $ global cost matrix $D$ is created. In this matrix, each element \( d(i,j) \) indicates the distance between two sub-sequences $\{x_1, x_2, ..., x_i\}$ and $\{y_1, y_2, ..., y_j\}$. The \( c(i,j) \) represents the cost (e.g., Euclidean distance) between elements \( x_i \) and \( y_j \). The alignment process begins with the pair (\( x_1 \), \( y_1 \)) and ends with the pair (\( x_n \), \( y_m \)). The first element \( d(1,1) \) is initialized with the distance between the initial elements of both sequences \( c(1,1) \), while the other elements are set to infinity. The matrix is then computed iteratively using the following formula:

\begin{equation}
d(i, j) = c(i, j) + \min \left\{ d(i-1,j), d(i,j-1), d(i-1,j-1) \right\} \
\end{equation}
The optimal path is determined by tracing back from \( d(n,m) \) to \( d(1,1) \), following the minimum value path. 


DTW is capable of recognizing similarities in the shapes of time-series data, even when there are variations in their length and speed. This aligns well with the tasks of human-AI collaborative writing. Humans might vary in the speed of essay writing, yet exhibit comparable patterns in AI-usage behaviors. For instance, some writers might prefer leveraging AI during the initial stages of writing, though the duration of this phase can differ.

\smallskip
\noindent\textbf{Clustering Method.} DTW functions solely as a metric to measure the similarity between times-series data of varying lengths; an additional clustering method to identify common groups of data instances is needed. In our case, we opted for K-means for clustering purposes, a widely recognized method for clustering time-series data \parencite{aghabozorgi2015time}. The optimal number of clusters was determined using the Elbow method \parencite{bholowalia2014ebk}, based on the DTW-inertia metric, which calculates the sum of the DTW distances between the data points and the corresponding centroid.

\smallskip 
\noindent\textbf{Feature Engineering.} We developed features to serve as inputs for DTW to assess the similarity between two writing sessions. Consistent with prior research that analyzed the writing process through keystroke dynamics \parencite{allen2016enter}, we initially segmented each writing session into discrete, non-overlapping 60-second time windows. For each of these time windows, we computed four AI-usage features, drawing from previous studies \parencite{shibani2023visual}. Consequently, this approach yielded time-series data with varying time steps, ranging from 2 to 32, and an average count of 11. We standardized features, adjusting their means to 0 and standard deviations to 1.0. Experiments were carried out using the tslearn toolkit \parencite{JMLR:v21:20-091}. The features are as follows:

\begin{itemize}
    \item \textbf{Number of GPT-3 Calls Made}: Number of seeking suggestions from GAI by human.
    \item \textbf{Rate of GPT-3 Suggestions Accepted}: Number of suggestions accepted by human / Number of seeking suggestions from GAI by human.
    \item \textbf{Rate of GPT-3 Suggestions Modified}: Number of suggestions modified by human / Number of suggestions accepted by human.
    \item \textbf{Rate of GPT-3 Generated Characters}: Number of characters completely authored by GAI / Total number of characters.
\end{itemize}

\smallskip
\noindent\textbf{Evaluation}: We initially conducted a Shapiro-Wilk Test and found that the data did not follow a normal distribution. Consequently, we opted for the non-parametric Mann-Whitney U test to evaluate differences between any two cluster pairings. To represent the temporal patterns of each cluster, we selected the cluster barycenter time-series, mapped into the longest time steps present in the original dataset, in our case, 32 time windows. This cluster barycenter time-series minimizes the DTW distances within its cluster, thus reflecting the overall temporal patterns of that cluster. Additionally, we included survey data to aid in interpreting the results in each cluster.

\subsection{RQ2: Epistemic Network Analysis}
ENA has demonstrated its effectiveness and validity as a method for assessing human-AI collaborative writing in previous research \parencite{cheng2024evidencecentered}. We adopted the evidence-centered assessment framework proposed in \parencite{cheng2024evidencecentered} (as outlined in Section \ref{sec:relatedwork}) to explore the relationships between temporal patterns of AI-usage and human behaviors specific to \texttt{knowledge telling}, \texttt{knowledge transformation}, and \texttt{cognitive presence} in writing processes. Specifically, ENA is a method used for quantifying and analyzing the structure of connections in coded data \parencite{shaffer2017epistemic}. This method contains four key concepts:
\begin{itemize}
    \item \textbf{Units}: Units are the subjects of analysis, with each one treated as an separate entity for which network data is generated and analyzed. In the current study, these units were differentiated based on writing session ID as recorded in the dataset we used.
    \item \textbf{Conversations}: Conversations refer to the sequence of interactions that are subject to analysis within each unit. In the current study, conversations were organized by session ID and sentence, facilitating the accumulation of co-occurrence.
    \item \textbf{Window Size}: Window Size refers to the number of coded elements that are considered together when building networks. In our study, we used an infinite stanza window, which computes the co-occurrences of each line in a conversation relative to every line that comes before it within the conversation.
    \item \textbf{Codes}: This involves categorizing segments of data into pre-established categories for analysis. Building on prior research on assessing human-AI collaborative writing \parencite{cheng2024evidencecentered}, the codes are presented in Table \ref{table1}. Certain writing behaviors are highly relevant to the framework's concepts. For example, the behaviors identified as \texttt{seekSugg} and \texttt{hoverSugg} correspond directly to specific stages of \texttt{cognitive presence}: \texttt{seekSugg} is associated with the \texttt{triggering event}, initiating the cognitive process, while \texttt{hoverSugg} relates to the \texttt{exploration} stage, where deeper engagement and understanding begin to develop. Conversely, some codes are associated with the framework based on whether connections exist between them. For instance, the link between \texttt{acceptSugg} and \texttt{compose} might suggest \texttt{knowledge telling}, whereas the concurrent presence of \texttt{acceptSugg}, \texttt{reviseSugg}, and \texttt{highModification} could imply \texttt{knowledge transformation}. Further details about code identification are provided in \parencite{cheng2024evidencecentered}.
\end{itemize}

ENA uses singular value decomposition (SVD) to map networks into a low-dimensional space. This process involves aligning the network graphs within the embedding space, and positioning the nodes and their connections to correspond with the most significant co-occurrences in each dimension. In these networks, the thickness of edge signifies the frequency of co-occurrence between two nodes, indicating the strength of connection. To further enhance the visualization of differences among clusters, we constructed a subtracted network \parencite{shaffer2016tutorial} between each pair of clusters. The color of each line identifies which of the two compared networks exhibits the stronger connection. A detailed explanation of ENA can be found in \parencite{bowman2021mathematical}. Our experiments were carried out using the ENA implementation in rENA \parencite{rENA}.

\begin{table*}[hbt!]
\begin{center}
\caption{Qualitative codes and definitions.}
\label{table1}
\resizebox{1\textwidth}{!}{
\begin{tabular}{@{}lll@{}}
\toprule
\textbf{Code} & \textbf{Definition} & \textbf{Identifiers} \\ \midrule
\texttt{compose} & Creating new content derived from the end of the existing text & Event name is "text-insert" in the end of text (space removed where applicable) \\ \midrule
\texttt{relocate} & Rearranging the sentences & The index of same sentence changes in last-current document \\ \midrule
\texttt{reflect} & Revising the content after or near completing the draft & Revise the content in the stage of at least finishing 90\% content (conclusion part-rule of thumb) \\ \midrule
\texttt{seekSugg} & Obtaining the suggestion & Event name is 'suggestion-get' \\ \midrule
\texttt{dismissSugg} & Dismissing the suggestion & Event name is 'suggestion-close' and event source us 'user' \\ \midrule
\texttt{acceptSugg} & Accepting the suggestion & Event name is 'suggestion-accept' \\ \midrule
\texttt{HoverSugg} & Hovering over the suggestions & Event name is 'suggestion-hover' \\ \midrule
\texttt{cursorFwd} & Moving the position of cursor forward & Event name is 'cursor-forward' \\ \midrule
\texttt{cursorBwd} & Moving the position of cursor backward & Event name is 'cursor-backward' \\ \midrule
\texttt{cursorSelect} & Selecting the text & Event name is 'cursor-select' \\ \midrule
\texttt{reviseUser} & Revising content they wrote & The inserts or deletes in-text content and ownership of revising sentence is 'user' \\ \midrule
\texttt{reviseSugg} & Revising the suggestion & The user inserts or deletes in-text content and owner- ship of revising sentence is 'api' \\ \midrule
\texttt{lowModification} & Making minor adjustments without altering the core meaning & Sentence semantic similarity \textgreater 0.8 \\ \midrule
\texttt{highModification} & Implementing significant changes to its meaning & Sentence semantic similarity \textless 0.8 \\ \bottomrule
\end{tabular}
}
\end{center}
\end{table*}

\begin{figure*}[hbt!]
\includegraphics[width=1\textwidth]{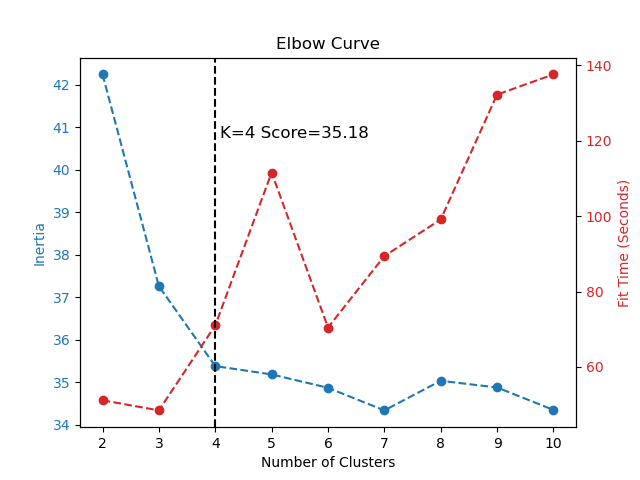}
\caption{The Elbow method plot.} 
\label{fig2}
\end{figure*}

\section{Results}
\subsection{Results on RQ1}

\begin{table*}[hbt!]
\begin{center}
\caption{Means and standard deviations of AI-usage features for each cluster.}
\label{table2}
\resizebox{1\textwidth}{!}{
\begin{tabular}{@{}lrrrr@{}}
\toprule
\multicolumn{1}{c}{\multirow{2}{*}{\textbf{Metrics}}} & \multicolumn{4}{c}{\textbf{Mean (Standard Deviation)}} \\ \cmidrule(l){2-5} 
\multicolumn{1}{c}{} & \textbf{\begin{tabular}[c]{@{}c@{}}Cluster 1  \\ (n = 168)\end{tabular}} & \textbf{\begin{tabular}[c]{@{}c@{}}Cluster 2 \\ (n = 368)\end{tabular}} & \textbf{\begin{tabular}[c]{@{}c@{}}Cluster 3 \\ (n = 550)\end{tabular}} & \textbf{\begin{tabular}[c]{@{}c@{}}Cluster 4 \\ (n = 359)\end{tabular}} \\ \midrule
\textbf{Number of GPT-3 Calls Made} & 10.07 (6.77) & 18.27 (12.10) & 7.86 (4.98) & 14.91 (7.53) \\
\textbf{Rate of GPT-3 Suggestions Accepted} & 0.62 (0.22) & 0.66 (0.31) & 0.65 (0.25) & 0.77 (0.16) \\
\textbf{Rate of GPT-3 Suggestions Modified} & 0.46 (0.30) & 0.17 (0.22) & 0.36 (0.31) & 0.34 (0.25) \\
\textbf{Rate of GPT-3 Generated Characters} & 0.20 (0.12) & 0.35 (0.21) & 0.18 (0.12) & 0.36 (0.17) \\ \bottomrule
\end{tabular}
}
\end{center}
\end{table*}

\begin{table*}[hbt!]
\begin{center}
\caption{Mann-Whitney U test of AI-usage features for pairs of clusters, asterisks indicate statistical significance ($****$: p$\leq$.0001; $***$: p$\leq$.001; $**$: p$\leq$.01; $*$: p$\leq$.05).}
\label{table3}
\resizebox{1\textwidth}{!}{
\begin{tabular}{@{}cS[table-format=6.1]S[table-format=6.1]S[table-format=6.1]S[table-format=6.1]@{}}
\toprule
\textbf{\begin{tabular}[c]{@{}c@{}}Compared \\ Clusters\end{tabular}} & \textbf{\begin{tabular}[c]{@{}c@{}}Number of GPT-3 \\ Calls Made\end{tabular}} & \textbf{\begin{tabular}[c]{@{}c@{}}Rate of GPT-3 \\ Suggestions Accepted\end{tabular}} & \textbf{\begin{tabular}[c]{@{}c@{}}Rate of GPT-3 \\ Suggestions Modified\end{tabular}} & \textbf{\begin{tabular}[c]{@{}c@{}}Rate of GPT-3 \\ Generated Characters\end{tabular}} \\ \midrule
\textbf{1 vs. 2} & 15613.0\fourS & 24531.5\threeS & 48499.5\fourS & 15930.0\fourS \\
\textbf{1 vs. 3} & 8104.5\fourS & 43799.0 & 55176.5\threeS & 52805.0\twoS \\
\textbf{1 vs. 4} & 15445.0\fourS & 18219.5\fourS & 36802.0\fourS & 12109.0\fourS \\
\textbf{2 vs. 3} & 159604.5\fourS & 111704.0\twoS & 69337.0\fourS & 154772.0\fourS \\
\textbf{2 vs. 4} & 79696.5\fourS & 57811.5\twoS & 36455.5\fourS & 71777.5\oneS \\
\textbf{3 vs. 4} & 35250.5\fourS & 70419.5\fourS & 96585.0 & 31773.0\fourS \\ \bottomrule
\end{tabular}
}
\end{center}
\end{table*}

\begin{table*}[hbt!]
\begin{center}
\caption{Means and standard deviations of survey data for each cluster.}
\label{table4}
\resizebox{1\textwidth}{!}{
\begin{tabular}{@{}lllll@{}}
\toprule
\multicolumn{1}{c}{\multirow{2}{*}{\textbf{Survey Questions}}} & \multicolumn{4}{c}{\textbf{Mean (Standard Deviation)}} \\ \cmidrule(l){2-5} 
\multicolumn{1}{c}{} & \textbf{Cluster 1} & \textbf{Cluster 2} & \textbf{Cluster 3} & \textbf{Cluster 4} \\ \midrule
\begin{tabular}[c]{@{}l@{}}Q1: During this writing session, the suggestions you received were grammatically \\ correct, and contributed to the fluency of the story/essay.\end{tabular} & 5.99 (1.22) & 5.89 (1.07) & 5.85 (1.24) & 6.09 (1.05) \\ \midrule
Q2: During this writing session, the suggestions helped me come up with new ideas. & 5.21 (1.79) & 5.24 (1.88) & 5.31 (1.81) & 5.57 (1.71) \\ \midrule
\begin{tabular}[c]{@{}l@{}}Q3: During this writing session, I felt like that I would have written a *better* \\ story/essay if I wrote the story alone.\end{tabular} & 3.01 (2.00) & 3.56 (2.01) & 3.00 (1.88) & 2.71 (1.85) \\ \midrule
\begin{tabular}[c]{@{}l@{}}Q4: During this writing session, the system was competent (having expert knowledge \\ and ability to perform a task successfully) in writing.\end{tabular} & 5.53 (1.56) & 5.44 (1.59) & 5.39 (1.64) & 5.89 (1.34) \\ \midrule
Q5: During this writing session, the system was capable of writing creative stories/essays. & 5.30 (1.71) & 5.33 (1.74) & 5.28 (1.75) & 5.65 (1.59) \\ \midrule
Q6: During this writing session, the system understood what I was trying to write. & 5.27 (1.70) & 5.63 (1.46) & 5.35 (1.69) & 5.72 (1.35) \\ \midrule
Q7: It was easy to write with the system. & 5.66 (1.66) & 5.70 (1.42) & 5.64 (1.60) & 6.02 (1.33) \\ \midrule
Q8: I am satisfied with the story/essay I wrote. & 5.83 (1.35) & 5.51 (1.53) & 5.81 (1.34) & 5.83 (1.50) \\ \midrule
Q9: I am confident in my ability to write a story/essay with the help of the system. & 5.87 (1.53) & 5.92 (1.31) & 5.90 (1.43) & 6.18 (1.22) \\ \bottomrule
\end{tabular}
}
\end{center}
\end{table*}

\begin{table*}[hbt!]
\begin{center}
\caption{Mann-Whitney U test of survey data for pairs of clusters, asterisks indicate statistical significance ($****$: p$\leq$.0001; $***$: p$\leq$.001; $**$: p$\leq$.01; $*$: p$\leq$.05).}
\label{table5}
\resizebox{1\textwidth}{!}{
\fontsize{8pt}{9pt}
\begin{tabular}{@{}cS[table-format=6.1]S[table-format=6.1]S[table-format=6.1]S[table-format=6.1]S[table-format=6.1]S[table-format=6.1]S[table-format=6.1]S[table-format=6.1]S[table-format=6.1]@{}}
\toprule
\textbf{\begin{tabular}[c]{@{}c@{}}Compared \\ Clusters\end{tabular}} & \textbf{Q1} & \textbf{Q2} & \textbf{Q3} & \textbf{Q4} & \textbf{Q5} & \textbf{Q6} & \textbf{Q7} & \textbf{Q8} & \textbf{Q9} \\ \midrule
\textbf{1 vs. 2} & 25284.0 & 22638.5 & 19432.0\twoS & 23765.0 & 22612.5 & 20591.5 & 23837.5 & 25758.5\oneS & 23904.0 \\
\textbf{1 vs. 3} & 34097.0 & 30544.0 & 31372.5 & 32927.0 & 31623.0 & 30654.0 & 32375.0 & 32151.0 & 32170.5 \\
\textbf{1 vs. 4} & 20426.5 & 18421.0\oneS & 22554.0 & 18195.0\oneS & 18241.0\oneS & 17988.0\oneS & 18627.0 & 20307.5 & 19005.0 \\
\textbf{2 vs. 3} & 78402.0 & 78876.5 & 92980.0\fourS & 81098.5 & 81850.5 & 86190.5 & 79437.5 & 71977.5\threeS & 78568.0 \\
\textbf{2 vs. 4} & 46310.5\twoS & 47711.5\oneS & 65767.0\fourS & 44597.5\threeS & 47537.5\oneS & 51347.0 & 45112.5\threeS & 45445.5\twoS & 45755.5\twoS \\
\textbf{3 vs. 4} & 64867.0\oneS & 66213.0\oneS & 79017.5\oneS & 60022.0\fourS & 63204.5\twoS & 64661.5\twoS & 62582.5\threeS & 68971.0 & 64186.5\twoS \\ \bottomrule
\end{tabular}
}
\end{center}
\end{table*}

\begin{figure*}[hbt!]
\includegraphics[width=1\textwidth]{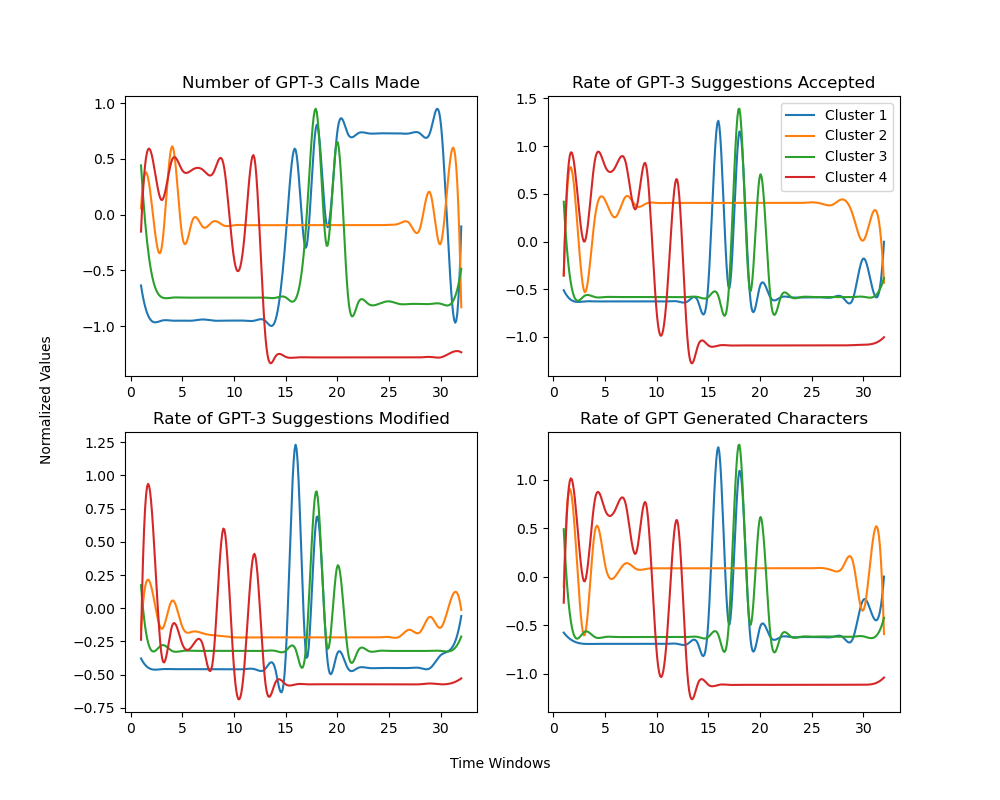}
\caption{The AI-usage features of time-series center for each cluster.} 
\label{fig3}
\end{figure*}

Our results in Figure \ref{fig2} show that the ideal number of clusters is 4. Our response to RQ1 was informed by examining both the overall AI-usage features in final essays and the temporal patterns of these AI-usage features. Additionally, we incorporated survey data for each cluster to enhance our interpretation of the results. The mean and standard deviation of AI-usage features different various clusters are presented in Table \ref{table2}. The Mann-Whitney U test results for the AI-usage features are provided in Table \ref{table3}. The mean and standard deviation of survey data across different clusters are outlined in Table \ref{table4}, and the results of the Mann-Whitney U test for survey data are displayed in Table \ref{table5}. The survey questions are also available in Table \ref{table4}. Figure \ref{fig3} displays the temporal patterns of clusters for four AI-usage features. After analyzing the results, we interpret our four clusters as follows:

\noindent\textbf{Cluster 1}: Writers in this group were generally less inclined to seek GAI suggestions in the initial stages of their writing process. However, they progressively relied more on GAI for providing suggestions in the latter half. These writers tended to accept and modify GAI suggestions more frequently during the mid-stages of their work, but did so less at the beginning and end. Consequently, AI-generated content was more prevalent in the middle of their writing, with fewer appearing at the start and finish. Notably, while these authors sought GAI suggestions more in the second half of their writing, they showed a reluctance to accept and modify these suggestions. This might indicate that writers sought AI suggestions, but did not receive ones that completely matched their expectations. Regarding the overall features of their final works, they demonstrated a moderate willingness to seek GAI suggestions. However, they had the lowest acceptance rate combined with the highest modification rate, leading to the second-lowest presence of AI-generated content. This pattern suggests that writers within this cluster showed a tendency towards limited use of GAI and were inclined to modify the suggestions given by GAI. The findings align with the survey data. Writers in Cluster 1 report the lowest average scores for Q2 and Q6, suggesting that they find the suggestions from the AI less helpful for generating new ideas and understanding their writing intentions compared to those in other clusters. Furthermore, the lowest average score for Q9 indicates that writers in Cluster 1 perceive it to be more challenging to improve their writing with AI's help.

\noindent \textbf{Cluster 2}: Writers within this cluster exhibited consistent behavior throughout their writing process. They consistently sought and accepted GAI suggestions at a relatively stable frequency, and also maintained a low rate of modification from start to finish. Consequently, the rate of AI-generated content remained nearly constant across all stages of their work. Regarding their final products, these writers showed the highest tendency to seek GAI suggestions, averaging over 18 times, and displayed the lowest modification rate (averaging 17\%). They also had the second-highest acceptance rate and the rate of AI-generated characters. This group can be characterized as having a high level of trust in GAI and a willingness to integrate it extensively into their writing. The survey data also reveal similar patterns. Writers in this group have the second-highest average scores for Q5, Q6, and Q9, suggesting they find AI assistance more beneficial and easier to use for writing compared to those in Clusters 1 and 3. Surprisingly, despite their extensive use of the AI in their writing, they have the highest score for Q3, suggesting they believe they can also write better without the AI's assistance.

\noindent \textbf{Cluster 3}: Writers in this cluster displayed temporal patterns in accepting suggestions, modifying suggestions, and rate of AI-generated content that was similar to that in Cluster 1, with lower levels at the beginning and end, and higher in the middle. However, unlike writers in Cluster 1 who continued to seek GAI suggestions in the latter half, writers in Cluster 3 showed a decreased inclination to seek GAI suggestions during the final phases. This suggests that they were likely not satisfied with the AI suggestions received during the middle phase and chose to rely less on AI assistance in the remainder of their writing. In terms of their final output, they had the lowest frequency of seeking GAI suggestions (averaging less than 8 times) and the lowest rate of AI-generated content (averaging 18\%), along with the second-highest modification rate. While writers in Clusters 1 and 3 might find that GAI suggestions did not meet their expectations, their approaches to this challenge varied. Cluster 1 tended to modify these suggestions, while Cluster 3 tended to minimize the use of GAI. Writers in Cluster 3 can be characterized as having less trust in GAI and a preference for relying on their own writing abilities. The survey data also corroborate our findings. Writers in this group have the lowest average scores for Q4 and Q5, indicating they share similar sentiments with writers in Cluster 1. Both groups are less confident about the effectiveness of GAI suggestions in enhancing their writing.

\noindent \textbf{Cluster 4}: Writers in this cluster predominantly sought GAI suggestions in the first half of their writing process and were less inclined to do so in the latter half. Similarly, they exhibited high acceptance and modification rates of these suggestions initially, which decreased in the second half. This pattern is also reflected in the presence of AI-generated characters, which was higher in the early stages and reduced later on. The behaviors of these writers suggest that they used GAI primarily for inspiration in the beginning stages, and once they gathered enough ideas, they tended to continue writing independently. Regarding their final works, these writers had the second-highest frequency of seeking GAI suggestions and the highest acceptance rate (averaging about 77\%). They also demonstrated the highest presence of AI-generated characters (averaging 36\%) and maintained a moderate level of modification. Writers in Cluster 4 can be described as striking a balance between using GAI and relying on their own writing skills. The survey data also validate our interpretation. Writers in this group have the highest average scores for Q2, Q4, Q5, Q6, and Q9, indicating a strong belief that GAI suggestions can enhance their writing and introduce new ideas. The lowest average score for Q3 further suggests that these writers feel significantly assisted by AI in their writing tasks.

When analyzing statistical tests, writers across different clusters exhibit significant differences regarding most AI-usage features in their final essays. Concerning survey data, there are no significant differences between Cluster 1 and Cluster 3, as detailed in Table \ref{table5}. This outcome is consistent with our observation that writers in both Cluster 1 and 3 might have found that GAI suggestions did not meet their expectations, although their approaches to this challenge differed. Nonetheless, significant differences are still evident between other pairs of clusters in certain survey questions, particularly when compared to Cluster 4.

\subsection{Results on RQ2}
\begin{figure*}[hbt!]
\centering
\includegraphics[width=1\textwidth]{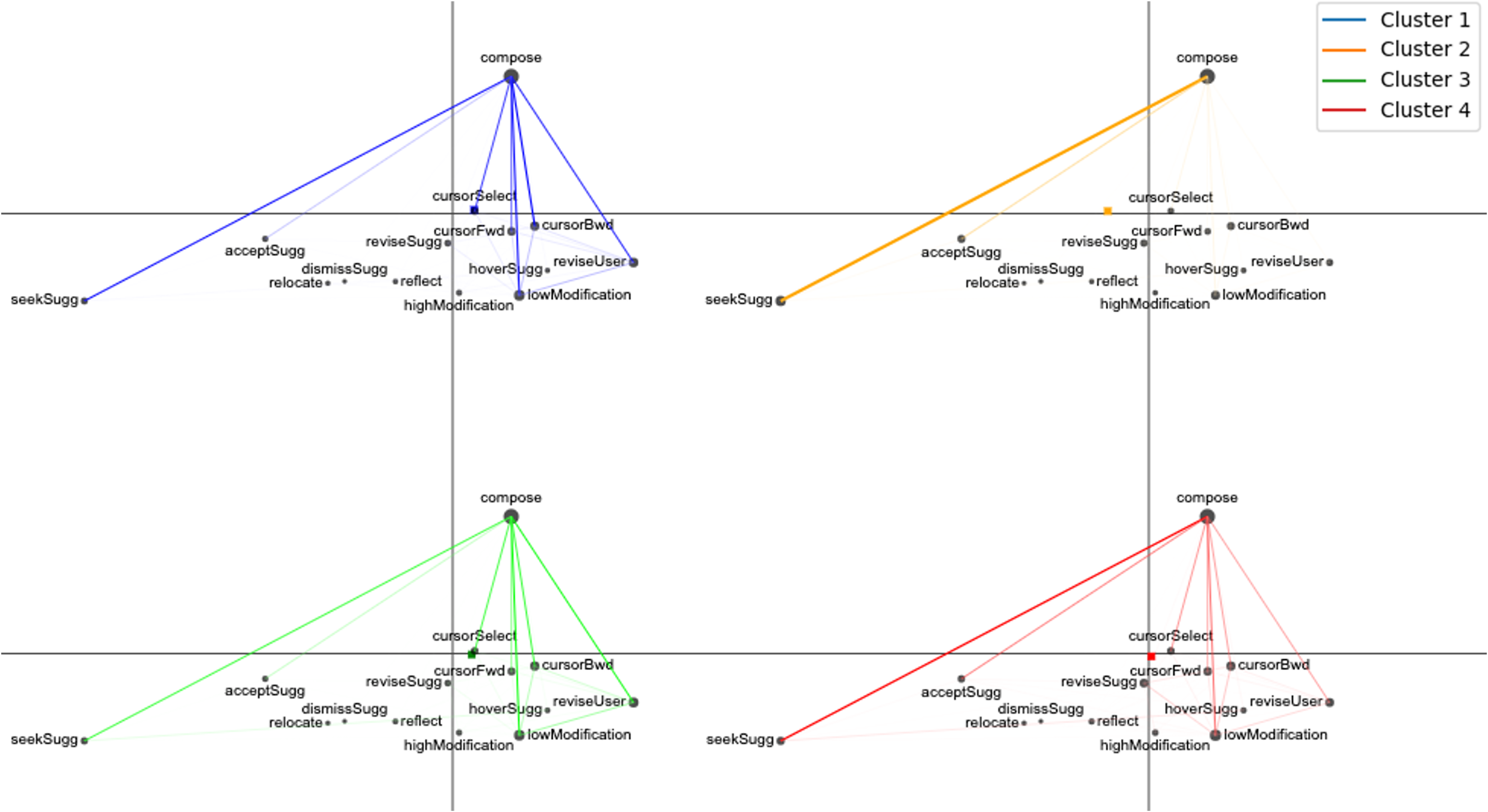}
\caption{ENA results for the four identified clusters.} 
\label{fig4}
\end{figure*}

\begin{figure*}[hbt!]
\centering
\includegraphics[width=1\textwidth]{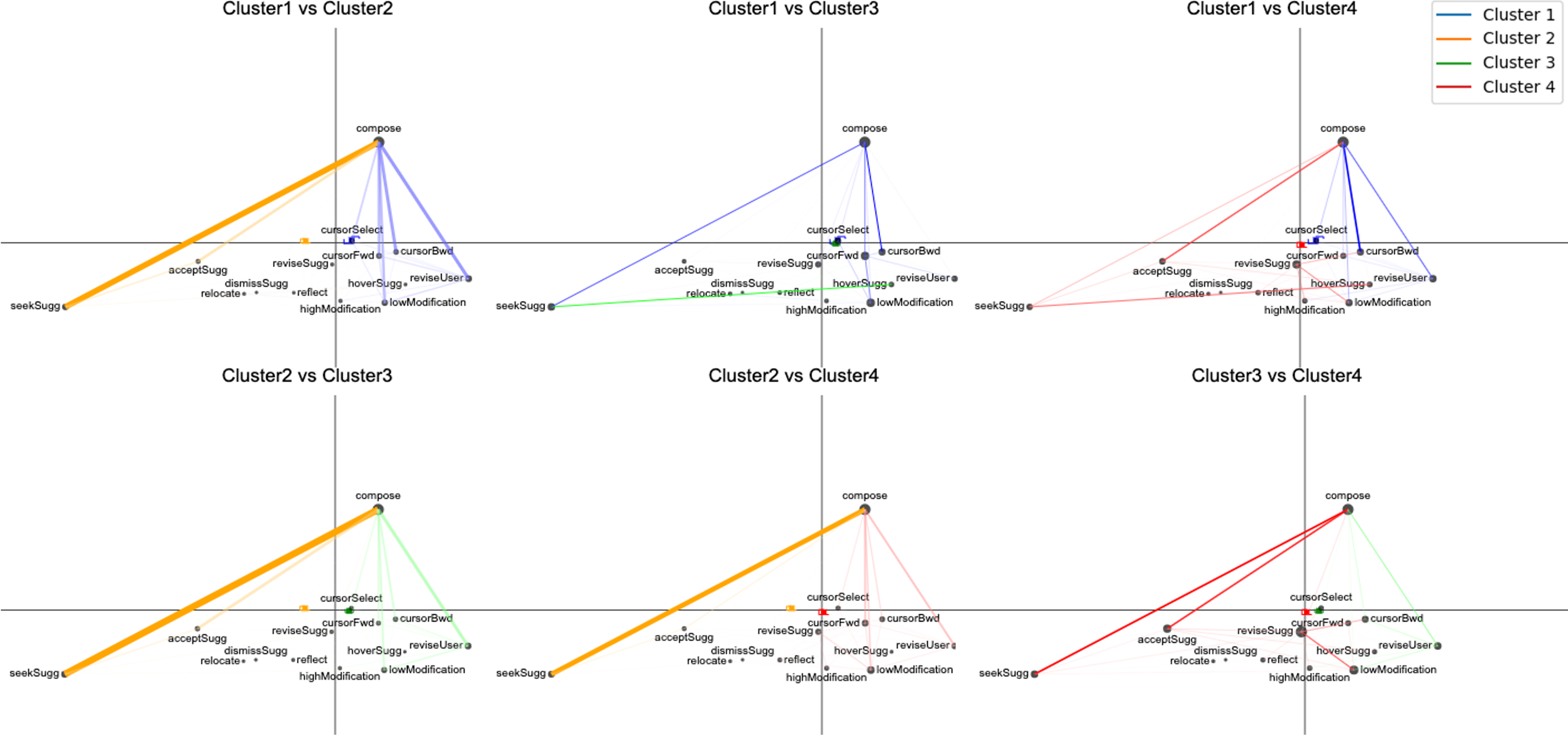}
\caption{Subtracted networks between pairs of clusters.} 
\label{fig5}
\end{figure*}

Results of ENA for each cluster are presented in Figure \ref{fig4}, while the subtracted networks for cluster pairs are depicted in Figure \ref{fig5}. Integrating the results from RQ1 with the ENA findings, several intriguing patterns emerge.

Cluster 1 demonstrated notably stronger connections between \texttt{compose} and \texttt{seekSugg}, and between \texttt{cursorFwd} and \texttt{lowModification} compared to Cluster 3. This aligns with our RQ1 observations, where both Clusters 1 and 3 showed a lack of confidence in GAI suggestions. However, Cluster 1 tended to modify these suggestions, while Cluster 3 leaned towards limited use of GAI. Moreover, Cluster 1 exhibited stronger connections between \texttt{compose} and \texttt{cursorBwd}, \texttt{reviseUser}, and \texttt{lowModification} compared to Clusters 2 and 4, indicating that writers in Cluster 1 were more inclined to revise and modify their essays. Consequently, writers in Cluster 1 showed an inclination towards the \texttt{integration} and \texttt{resolution} facets of \texttt{cognitive presence}. They preferred to synthesize suggestions from AI, aiming to produce a cohesive piece of writing. 

Cluster 2 distinguished itself with the strongest connection between \texttt{compose} and both \texttt{seekSugg} and \texttt{acceptSugg} among all clusters, aligning with RQ1 findings that writers in this cluster had considerable trust in GAI and were more inclined to incorporate it extensively in their writing. Consequently, writers in Cluster 2 predominantly employed the \texttt{knowledge telling} strategy. They accepted AI-generated suggestions as they were, essentially adhering strictly to the guidance provided by the AI without making any modifications.

Cluster 3 was distinguished by the weakest connection between \texttt{compose} and \texttt{seekSugg} among clusters. This suggests the lowest trust in GAI. Additionally, Cluster 3 exhibited stronger connections between \texttt{seekSugg} and \texttt{hoverSugg} compared to Cluster 1, indicating that writers in Cluster 3 frequently hesitated among GAI suggestions before ultimately deciding not to accept them. Consequently, writers in Cluster 3 exhibited an inclination for the \texttt{exploration} aspect of \texttt{cognitive presence}. They interacted with AI to evaluate various suggestions. 

Cluster 4 showed stronger connections between \texttt{compose} and \texttt{seekSug}, between \texttt{compose} and \texttt{acceptSug}, and between \texttt{reviseSugg} and \texttt{lowModification} compared to Clusters 1 and 3. Consistent with RQ1 insights, writers in Cluster 4 maintained a balance between using GAI and relying on their writing skills, showing a greater openness to integrating GAI into their writing than those in Clusters 1 and 3. Additionally, in comparison to Cluster 2, Cluster 4 displayed stronger connections between \texttt{reviseSugg} and \texttt{lowModification}. This suggests that, unlike writers in Cluster 2 who completely trusted AI, those in Cluster 4 preferred to blend AI suggestions with their own knowledge. Consequently, writers in cluster 4 demonstrated a preference for employing the \texttt{knowledge transformation} strategy. They typically initiated their revision process by utilizing AI-generated suggestions as a foundation, which they then customized to align with their requirements.

\section{Discussion and Conclusion}
In this study, we investigated common temporal patterns of GAI usage in human-AI collaborative writing, along with the effects of GAI usage on human writing behaviors that reflect cognitive processes. Specifically, we applied the DTW time-series clustering method to identify distinct clusters and interpreted these clusters based on the AI-usage features, considering both temporal patterns and the final products. 
Additionally, we incorporated survey data to enhance our understanding of the patterns in human-AI collaborative writing processes. To delve deeper into the connection between temporal patterns of AI usage and cognitive processes in writing, we conducted ENA for each cluster and made comparisons between pairs of clusters. We identified four common temporal patterns of GAI usage in human-AI collaborative writing, and described their relationships to cognitive processes: Cluster 1 indicates low trust in AI and willingness to modify AI suggestions, correlating with the \texttt{integration} and \texttt{resolution} aspects of \texttt{cognitive presence} \parencite{garrison1999critical}; Cluster 2 demonstrates a high intention to use AI and extensive incorporation of AI-generated suggestions during the whole writing process, adopting the \texttt{knowledge telling} strategy \parencite{bereiter2013psychology}; Cluster 3 exhibits low trust in AI with an inclination for independent writing, which aligns with the \texttt{exploration} aspect of \texttt{cognitive presence} \parencite{garrison1999critical}; and Cluster 4 presents a balanced approach to AI usage and self-writing, implementing the \texttt{knowledge transformation} strategy \parencite{bereiter2013psychology}.

Based on identified common patterns in human-AI collaborative writing, we argue that those patterns share similarities with patterns previously identified in traditional, human-only collaborative writing tasks in terms of collaborative models \parencite{zhang2009designs}, writing approaches \parencite{heinonen2020scripting}, and human profiles during collaborative writing \parencite{marttunen2012participant}. In this context, GAI acts as non-human collaborators, working alongside human writers. Firstly, we suggest that collaborative models between GAI and humans are somewhat similar to peer collaboration models \parencite{zhang2009designs}. There are three types of peer collaboration models: \texttt{fixed peer collaboration model}, \texttt{interactive peer collaboration model}, and \texttt{opportunistic peer collaboration model}. In the \texttt{fixed peer collaboration model}, team members have defined roles, tasks, and responsibilities that remain constant throughout a project. Similarly, in human-AI collaboration, both parties may have distinct, predefined roles. For example, the AI could generate ideas, while the human focuses on composing subsequent paragraphs, making minimal changes to the AI's suggestions. Writers in Cluster 2 may prefer to follow this collaboration model. In the \texttt{interactive peer collaboration model}, team members exchange knowledge and engage with each other actively. In human-AI scenarios, the AI might suggest edits or content expansions based on its analysis of the text, and the human can accept, reject, or modify these suggestions. Writers in Cluster 4 may prefer this collaboration model. In the \texttt{opportunistic peer collaboration model}, team members have the flexibility to work independently and collaborate as needed. In human-AI collaboration, the human might write independently and consult the AI for input only when necessary. Writers in Clusters 1 and 3 may prefer to follow this collaboration model. 

Secondly, the identified human-AI collaborative writing patterns align with the human-only collaborative writing approaches proposed by \parencite{heinonen2020scripting}. In the \texttt{script based writing approach}, the script provides structure and clarity, outlining each group member's tasks at specific moments. This is similar to the approach used by writers in Cluster 2, where GAI and humans maintain constant predefined roles in the writing process. In contrast, the \texttt{collective writing approach} might be seen as more effective by some groups, who may find the script meaningless or difficult to follow and prefer a more flexible method. This resembles the approach used by writers in Cluster 4, where writers can use GAI to suggest texts based on their content or modify GAI suggestions to suit their needs, with interchangeable roles. Additionally, in the \texttt{separate writing approach}, writing tasks are divided into distinct parts among group members, rather than collaboratively producing a single document. Writers in Clusters 1 and 3 may partially prefer this approach, as they sometimes like to write independently without GAI assistance. However, GAI still needs to process the human-written text to generate suggestions, which means the human-written content can influence the GAI-written content, making the separate writing less distinct in human-AI collaborative writing tasks.

Thirdly, we argue that the human profiles observed during human-only collaborative writing tasks, as proposed by \parencite{marttunen2012participant}, share similarities with the identified clusters in the present study. In human-only collaborative writing tasks, \texttt{cognitively focused thinkers} refer to students who primarily focus on the writing task itself, spending most of their time discussing the topic, planning, writing, and revising the text. Writers in Cluster 1 exhibited similar behavior, i.e., they appeared to critically engage with the content and refine it to ensure accuracy and coherence. \texttt{Performance steering writers} are individuals who take charge of the actual writing and revising activities, often steering the group's performance and exchanging concepts less frequently. This aligns with writers in Cluster 3, who prefer to write more independently, with a limited exchange of ideas with GAI. \texttt{Textbook consulters} are students who frequently refer to the textbook while writing, ensuring their work closely aligns with theoretical ideas presented in the text and seldom presenting their own ideas. Writers in Cluster 2 show similarities with this profile, relying heavily on GAI suggestions and seldom modifying these suggestions. \texttt{Cognitively versatile thinkers} are students who engage in a wide range of cognitive activities and present their own ideas. Writers in Cluster 4 are similar to this profile, as they prefer to transform and integrate AI-generated content with their own knowledge, creating a coherent and well-informed text.

\subsection{Implications}
We argue that researchers and educators can develop and fine-tune GAI writing tools customized to the temporal patterns of GAI usage and the cognitive processes demonstrated by writers. For example, writers who show low trust in AI and willingness to modify AI suggestions should not receive explicit sentence-level suggestions from writing tools. Instead, these tools should offer conceptual ideas that writers can integrate into their essays. For writers displaying low trust in AI with a preference for independent writing, educators might focus on enhancing the quality of AI-generated suggestions to better align with the writers' expectations. For writers using AI to guide their initial brainstorming, AI-generated suggestions should be inspiring during this initial stage. As the essay progresses, the focus should shift towards enhancing the quality of the writing (e.g., coherence and lexical diversity). 

Additionally, given the similarities identified between human-AI collaborative writing and human-only collaborative writing in terms of collaborative models, writing approaches, and human profiles during collaborative writing, educators can adapt existing human-only collaborative writing approaches and theories to this cutting-edge human-AI context. This adaptation will help in designing effective and efficient human-AI collaborative writing tools that cater to human cognitive needs and collaborative model preferences. For instance, writers who prefer the \texttt{fixed peer collaboration model} would benefit from fixed functional and less flexible GAI-writing tools tailored to their needs. On the other hand, writers who prefer the \texttt{interactive peer collaboration model} would require more complex and multifunctional GAI-writing tools to meet a variety of different requirements.

\subsection{Limitations}
We acknowledge several limitations in our study. Firstly, CoAuthor provides a limited range of interactions with GAI, such as not permitting GAI to modify human-written content, which confines our ability to identify even more diverse patterns. Secondly, our findings are based on the CoAuthor dataset and may not be necessarily applicable to other writing contexts involving different subjects or requirements. Thirdly, our analysis did not differentiate data according to essay genres, such as creative vs. argumentative writing. Given that writing patterns can differ among genres, this area merits further research in future studies.


\printbibliography

@inproceedings{lee2022coauthor,
  title={Coauthor: Designing a human-ai collaborative writing dataset for exploring language model capabilities},
  author={Lee, Mina and Liang, Percy and Yang, Qian},
  booktitle={Proceedings of the 2022 CHI conference on human factors in computing systems},
  pages={1--19},
  year={2022}
}

@article{fareed2016esl,
  title={ESL learners’ writing skills: Problems, factors and suggestions},
  author={Fareed, Muhammad and Ashraf, Almas and Bilal, Muhammad},
  journal={Journal of education and social sciences},
  volume={4},
  number={2},
  pages={81--92},
  year={2016}
}

@article{dj2015using,
  title={Using hypnoteaching strategy to improve students’ writing ability},
  author={Dj, Muhammad Zuhri and Sukarnianti, Sukarnianti},
  journal={Dinamika Ilmu},
  volume={15},
  number={2},
  pages={185--199},
  year={2015}
}

@article{setyowati2016analyzing,
  title={Analyzing the students’ ability in writing opinion essay using flash fiction},
  author={Setyowati, Lestari},
  journal={Journal of English Language Teaching and Linguistics},
  volume={1},
  number={1},
  pages={79--92},
  year={2016}
}

@article{roe2023review,
  title={A Review of AI-Powered Writing Tools and Their Implications for Academic Integrity in the Language Classroom},
  author={Roe, Jasper and Renandya, Willy A and Jacobs, George M},
  journal={Journal of English and Applied Linguistics},
  volume={2},
  number={1},
  pages={3},
  year={2023}
}

@article{chang2023survey,
  title={A survey on evaluation of large language models},
  author={Chang, Yupeng and Wang, Xu and Wang, Jindong and Wu, Yuan and Zhu, Kaijie and Chen, Hao and Yang, Linyi and Yi, Xiaoyuan and Wang, Cunxiang and Wang, Yidong and others},
  journal={arXiv preprint arXiv:2307.03109},
  year={2023}
}

@article{zhao2022leveraging,
  title={Leveraging artificial intelligence (AI) technology for English writing: Introducing Wordtune as a digital writing assistant for EFL writers},
  author={Zhao, Xin},
  journal={RELC Journal},
  pages={00336882221094089},
  year={2022},
  publisher={SAGE Publications Sage UK: London, England}
}

@inproceedings{cheng2024evidencecentered,
  title={Evidence-centered Assessment for Writing with Generative AI},
  author={Cheng, Yixin and Lyons, Kayley and Chen, Guanliang and Ga{\v{s}}evic, Dragan and Swiecki, Zachari},
  booktitle={The 14th Learning Analytics and Knowledge Conference (LAK’24)},
  year={2024},
}

@misc{wambsganss2023unraveling,
      title={Unraveling Downstream Gender Bias from Large Language Models: A Study on AI Educational Writing Assistance}, 
      author={Thiemo Wambsganss and Xiaotian Su and Vinitra Swamy and Seyed Parsa Neshaei and Roman Rietsche and Tanja Käser},
      year={2023},
      eprint={2311.03311},
      archivePrefix={arXiv},
      primaryClass={cs.CL}
}

@article{coenen2021wordcraft,
  title={Wordcraft: a human-ai collaborative editor for story writing},
  author={Coenen, Andy and Davis, Luke and Ippolito, Daphne and Reif, Emily and Yuan, Ann},
  journal={arXiv preprint arXiv:2107.07430},
  year={2021}
}

@inproceedings{kreminski2020we,
  title={Why Are We Like This?: The AI architecture of a co-creative storytelling game},
  author={Kreminski, Max and Dickinson, Melanie and Mateas, Michael and Wardrip-Fruin, Noah},
  booktitle={Proceedings of the 15th International Conference on the Foundations of Digital Games},
  pages={1--4},
  year={2020}
}

@article{shibani2023visual,
  title={Visual Representation of Co-Authorship with GPT-3: Studying Human-Machine Interaction for Effective Writing.},
  author={Shibani, Antonette and Rajalakshmi, Ratnavel and Mattins, Faerie and Selvaraj, Srivarshan and Knight, Simon},
  journal={International Educational Data Mining Society},
  year={2023},
  publisher={ERIC}
}

@article{muller2007dynamic,
  title={Dynamic time warping},
  author={M{\"u}ller, Meinard},
  journal={Information retrieval for music and motion},
  pages={69--84},
  year={2007},
  publisher={Springer}
}

@article{shaffer2017epistemic,
  title={Epistemic network analysis: A worked example of theory-based learning analytics},
  author={Shaffer, D and Ruis, A},
  journal={Handbook of learning analytics},
  year={2017}
}

@article{rai2010survey,
  title={A survey of clustering techniques},
  author={Rai, Pradeep and Singh, Shubha},
  journal={International Journal of Computer Applications},
  volume={7},
  number={12},
  pages={1--5},
  year={2010},
  publisher={International Journal of Computer Applications, 244 5 th Avenue,\# 1526, New~…}
}

@article{aghabozorgi2015time,
  title={Time-series clustering--a decade review},
  author={Aghabozorgi, Saeed and Shirkhorshidi, Ali Seyed and Wah, Teh Ying},
  journal={Information systems},
  volume={53},
  pages={16--38},
  year={2015},
  publisher={Elsevier}
}

@article{talebinamvar2022clustering,
  title={Clustering students’ writing behaviors using keystroke logging: a learning analytic approach in EFL writing},
  author={Talebinamvar, Mobina and Zarrabi, Forooq},
  journal={Language Testing in Asia},
  volume={12},
  number={1},
  pages={6},
  year={2022},
  publisher={Springer}
}

@inproceedings{zhang2019identifying,
  title={Identifying and comparing writing process patterns using keystroke logs},
  author={Zhang, Mo and Zhu, Mengxiao and Deane, Paul and Guo, Hongwen},
  booktitle={Quantitative Psychology: 83rd Annual Meeting of the Psychometric Society, New York, NY 2018},
  pages={367--381},
  year={2019},
  organization={Springer}
}

@article{hung2015identifying,
  title={Identifying at-risk students for early interventions—A time-series clustering approach},
  author={Hung, Jui-Long and Wang, Morgan C and Wang, Shuyan and Abdelrasoul, Maha and Li, Yaohang and He, Wu},
  journal={IEEE Transactions on Emerging Topics in Computing},
  volume={5},
  number={1},
  pages={45--55},
  year={2015},
  publisher={IEEE}
}

@inproceedings{cobo2010modeling,
  title={Modeling students' activity in online discussion forums: a strategy based on time series and agglomerative hierarchical clustering},
  author={Cobo, Germ{\'a}n and Garc{\'\i}a-Sol{\'o}rzano, David and Santamar{\'\i}a, Eug{\`e}nia and Mor{\'a}n, Jose Antonio and Melench{\'o}n, Javier and Monzo, Carlos},
  booktitle={Educational data mining 2011},
  year={2010}
}

@article{hassan2022utilizing,
  title={Utilizing social clustering-based regression model for predicting student’s GPA},
  author={Hassan, Yomna MI and Elkorany, Abeer and Wassif, Khaled},
  journal={IEEE Access},
  volume={10},
  pages={48948--48963},
  year={2022},
  publisher={IEEE}
}

@article{shen2017clustering,
  title={Clustering Student Sequential Trajectories Using Dynamic Time Warping.},
  author={Shen, Shitian and Chi, Min},
  journal={International Educational Data Mining Society},
  year={2017},
  publisher={ERIC}
}

@inproceedings{ma2022grading,
  title={Grading Problem-Solving for Clustering Students' Score Using Dynamic Programming Procedure in The Context of Dynamic Time Warping},
  author={Ma'ady, Mochamad Nizar Palefi and Syahda, Tabina Shafa Nabila and Nasrullah, Muhammad and Salsabila, Anindya Salwa and Asfari, Ully and Mardhiana, Hawwin},
  booktitle={2022 Seventh ICIC},
  pages={1--5},
  year={2022},
  organization={IEEE}
}

@article{he2023clustering,
  title={Clustering sequential navigation patterns in multiple-source reading tasks with dynamic time warping method},
  author={He, Qiwei and Borgonovi, Francesca and Su{\'a}rez-{\'A}lvarez, Javier},
  journal={Journal of Computer Assisted Learning},
  volume={39},
  number={3},
  pages={719--736},
  year={2023},
  publisher={Wiley Online Library}
}

@article{revesz2017effects,
  title={Effects of task complexity on L2 writing behaviors and linguistic complexity},
  author={R{\'e}v{\'e}sz, Andrea and Kourtali, Nektaria-Efstathia and Mazgutova, Diana},
  journal={Language Learning},
  volume={67},
  number={1},
  pages={208--241},
  year={2017},
  publisher={Wiley Online Library}
}

@article{koppenhaver2010conceptual,
  title={A conceptual review of writing research in augmentative and alternative communication},
  author={Koppenhaver, David and Williams, Amy},
  journal={Augmentative and Alternative Communication},
  volume={26},
  number={3},
  pages={158--176},
  year={2010},
  publisher={Taylor \& Francis}
}

@article{anak2012relationship,
  title={The relationship between English writing ability levels and EFL learners’ metacognitive behavior in the writing process},
  author={anak Engkamat, Daphne Tunga and Nasri, Nurfaradilla Mohamad},
  journal={International Journal of Academic Research in Progressive Education and Development},
  year={2012}
}

@inproceedings{knowles2022human,
  title={Human-AI Collaborative Writing: Sharing the Rhetorical Task Load},
  author={Knowles, Alan M},
  booktitle={2022 IEEE International Professional Communication Conference (ProComm)},
  pages={257--261},
  year={2022},
  organization={IEEE}
}

@article{senin2008dynamic,
  title={Dynamic time warping algorithm review},
  author={Senin, Pavel},
  journal={University of Hawaii at Manoa Honolulu, USA},
  volume={855},
  number={1-23},
  pages={40},
  year={2008}
}

@article{JMLR:v21:20-091,
  author  = {Romain Tavenard and Johann Faouzi and Gilles Vandewiele and
             Felix Divo and Guillaume Androz and Chester Holtz and
             Marie Payne and Roman Yurchak and Marc Ru{\ss}wurm and
             Kushal Kolar and Eli Woods},
  title   = {Tslearn, A Machine Learning Toolkit for Time Series Data},
  journal = {Journal of Machine Learning Research},
  year    = {2020},
  volume  = {21},
  number  = {118},
  pages   = {1-6},
  url     = {http://jmlr.org/papers/v21/20-091.html}
}

@article{allen2016enter,
  title={$\{$ENTER$\}$ ing the Time Series $\{$SPACE$\}$: Uncovering the Writing Process through Keystroke Analyses.},
  author={Allen, Laura K and Jacovina, Matthew E and Dascalu, Mihai and Roscoe, Rod D and Kent, Kevin M and Likens, Aaron D and McNamara, Danielle S},
  journal={IEDMS},
  year={2016},
  publisher={ERIC}
}

@inproceedings{bowman2021mathematical,
  title={The mathematical foundations of epistemic network analysis},
  author={Bowman, Dale and Swiecki, Zachari and Cai, Zhiqiang and Wang, Yeyu and Eagan, Brendan and Linderoth, Jeff and Shaffer, David Williamson},
  booktitle={Advances in Quantitative Ethnography: Second International Conference, ICQE 2020, Malibu, CA, USA, February 1-3, 2021, Proceedings 2},
  pages={91--105},
  year={2021},
  organization={Springer}
}

@Manual{rENA,
   title = {rENA: Epistemic Network Analysis},
   author= {Cody L, Marquart and Zachari, Swiecki and Wesley, Collier and Brendan, Eagan and Roman, Woodward and David, Williamson},
   year={2022},
}

@article{shaffer2016tutorial,
  title={A tutorial on epistemic network analysis: Analyzing the structure of connections in cognitive, social, and interaction data},
  author={Shaffer, David Williamson and Collier, Wesley and Ruis, Andrew R},
  journal={Journal of Learning Analytics},
  volume={3},
  number={3},
  pages={9--45},
  year={2016}
}

@article{zhang2009designs,
  title={Designs for collective cognitive responsibility in knowledge-building communities},
  author={Zhang, Jianwei and Scardamalia, Marlene and Reeve, Richard and Messina, Richard},
  journal={The Journal of the learning sciences},
  volume={18},
  number={1},
  pages={7--44},
  year={2009},
  publisher={Taylor \& Francis}
}

@article{zhang2015process,
  title={Process features in writing: Internal structure and incremental value over product features},
  author={Zhang, Mo and Deane, Paul},
  journal={ETS Research Report Series},
  volume={2015},
  number={2},
  pages={1--12},
  year={2015},
  publisher={Wiley Online Library}
}

@article{sinharay2019prediction,
  title={Prediction of essay scores from writing process and product features using data mining methods},
  author={Sinharay, Sandip and Zhang, Mo and Deane, Paul},
  journal={Applied Measurement in Education},
  volume={32},
  number={2},
  pages={116--137},
  year={2019},
  publisher={Taylor \& Francis}
}

@article{leijten2013keystroke,
  title={Keystroke logging in writing research: Using Inputlog to analyze and visualize writing processes},
  author={Leijten, Mari{\"e}lle and Van Waes, Luuk},
  journal={Written Communication},
  volume={30},
  number={3},
  pages={358--392},
  year={2013},
  publisher={Sage Publications Sage CA: Los Angeles, CA}
}

@article{latif2021remodeling,
  title={Remodeling writers’ composing processes: Implications for writing assessment},
  author={Latif, Muhammad MM Abdel},
  journal={Assessing Writing},
  volume={50},
  pages={100547},
  year={2021},
  publisher={Elsevier}
}

@article{conijn2019understanding,
  title={Understanding the keystroke log: the effect of writing task on keystroke features},
  author={Conijn, Rianne and Roeser, Jens and Van Zaanen, Menno},
  journal={Reading and Writing},
  volume={32},
  number={9},
  pages={2353--2374},
  year={2019},
  publisher={Springer}
}

@article{baaijen2012keystroke,
  title={Keystroke analysis: Reflections on procedures and measures},
  author={Baaijen, Veerle M and Galbraith, David and De Glopper, Kees},
  journal={Written Communication},
  volume={29},
  number={3},
  pages={246--277},
  year={2012},
  publisher={Sage Publications Sage CA: Los Angeles, CA}
}

@article{nguyen2024human,
  title={Human-AI collaboration patterns in AI-assisted academic writing},
  author={Nguyen, Andy and Hong, Yvonne and Dang, Belle and Huang, Xiaoshan},
  journal={Studies in Higher Education},
  pages={1--18},
  year={2024},
  publisher={Taylor \& Francis}
}

@book{bereiter2013psychology,
  title={The psychology of written composition},
  author={Bereiter, Carl and Scardamalia, Marlene},
  year={2013},
  publisher={Routledge}
}

@article{garrison1999critical,
  title={Critical inquiry in a text-based environment: Computer conferencing in higher education},
  author={Garrison, D Randy and Anderson, Terry and Archer, Walter},
  journal={The internet and higher education},
  volume={2},
  number={2-3},
  pages={87--105},
  year={1999},
  publisher={Elsevier}
}

@article{Vakkari2021PredictingEQ,
  title={Predicting essay quality from search and writing behavior},
  author={Pertti Vakkari and Michael V{\"o}lske and Martin Potthast and Matthias Hagen and Benno Stein},
  journal={Journal of the Association for Information Science and Technology},
  year={2021},
  volume={72},
  pages={839 - 852},
  url={https://api.semanticscholar.org/CorpusID:234165403}
}

@article{bholowalia2014ebk,
  title={EBK-means: A clustering technique based on elbow method and k-means in WSN},
  author={Bholowalia, Purnima and Kumar, Arvind},
  journal={International Journal of Computer Applications},
  volume={105},
  number={9},
  year={2014},
  publisher={Citeseer}
}

@article{heinonen2020scripting,
  title={Scripting as a pedagogical method to guide collaborative writing: university students’ reflections},
  author={Heinonen, Kirsi and De Grez, Nore and H{\"a}m{\"a}l{\"a}inen, Raija and De Wever, Bram and van der Meijs, Sophie},
  journal={Research and Practice in Technology Enhanced Learning},
  volume={15},
  pages={1--20},
  year={2020},
  publisher={Springer}
}

@article{marttunen2012participant,
  title={Participant profiles during collaborative writing},
  author={Marttunen, Miika and Laurinen, Leena},
  journal={Journal of writing research},
  volume={4},
  number={1},
  pages={53--79},
  year={2012}
}
\end{document}